# Breast mass detection in digital mammography based on anchor-free architecture

Haichao Cao[1].

[1] Hangzhou Hikvision Digital Technology Co., Ltd., Zhejiang, 310051, China

*E-mails: dr_caohaichao@gmail.com

**Abstract**

Background and Objective: Accurate detection of breast masses in mammography images is critical to diagnose early breast cancer, which can greatly improve the patients' survival rate. However, it is still a big challenge due to the heterogeneity of breast masses and the complexity of their surrounding environment.

Methods: To address these problems, we propose a one-stage object detection architecture, called Breast Mass Detection Network (BMassDNet), based on anchor-free and feature pyramid which makes the detection of breast masses of different sizes well adapted. We introduce a truncation normalization method and combine it with adaptive histogram equalization to enhance the contrast between the breast mass and the surrounding environment. Meanwhile, to solve the overfitting problem caused by small data size, we propose a natural deformation data augmentation method and mend the train data dynamic updating method based on the data complexity to effectively utilize the limited data. Finally, we use transfer learning to assist the training process and to improve the robustness of the model ulteriorly.

Results: On the INbreast dataset, each image has an average of 0.495 false positives whilst the recall rate is 0.930; On the DDSM dataset, when each image has 0.599 false positives, the recall rate reaches 0.943.

Conclusions: The experimental results on datasets INbreast and DDSM show that the proposed BMassDNet can obtain competitive detection performance over the current top ranked methods.

# 1  Introduction

Breast cancer is currently the most common cancer in the global female population [1]. According to statistics, in 2018, the number of cancer patients worldwide exceeded 18.1 million, of which the number of breast cancer patients achieved 3.8 million, accounting for 11.6% in total [2]. Clinical experience has shown that the diagnosis of early breast cancer is critical for improving the patients' survival rate [3–5]. Because of the clear imaging and sensitivity to early breast masses, mammography images have become the preferred imaging diagnostic method [6]. However, current examination of mammography images is mainly dependent on the subjective experience of physicians, which would lead to the missed detection and misdetection due to visual fatigue and loss of attention [7–9]. To effectively avoid the cumbersome manual labeling and the variability of detection results, the development of a robust breast mass automatic detection system has important clinical significance [10].

However, the heterogeneity of the breast mass and its similarity to the visual characteristics of surrounding tissue, is very detrimental to the development of a robust detection model. As shown in Fig. 1, breast mass (BMass) presents a great differences in either size, shape, or intensity in different cases. For example, the BMass's size in Fig. 1(d) is very large, the BMass in Fig. 1(b) has large intensity, whilst the size and intensity of BMass in Fig. 1(a, e) are relatively small. Furthermore, from Fig. 1 (c, e), it can be found that the BMass and other tissues, as the nipple of Fig. 1 (c), or the background in Fig. 1 (e) have similar visual features. It should be noted that since the original mammogram is very large, in order to better observe the BMass region, only the rectangular region corresponding to the breast is shown here.

To solve the above problems, we conduct BMass detection based on the one-stage object detection architecture FSAF [11] proposed by Zhu et al. in 2019, which is called as the Breast Mass Detection Network (BMassDNet). Although BMassDNet does not make significant improvements in the network architecture, based on the characterization of BMass in the mammography image, we proposed a normalization method and a data augmentation method to improve the detection performance. Also, to make better use of limited training data, we have proposed a training method for data dynamic updates during training process. In general, BMassDNet can detect different types of BMass lesions and obtain excellent detection performance. Our technical contributions in this work can be concluded into the following four aspects.

(1) To the best of our knowledge, this paper is the first to apply FSAF structure to breast mass detection. Besides, we have further improved the FASF architecture through horizontal connection and upsampling.

(2) To alleviate the problem that the BMass is similar to the intensity of the surrounding tissue, we propose a new normalization method and image enhancement method. They are combined to stretch the intensity on original image so as to make the BMass border clearer.

(3) Due to the limitation of the size of data set, it is difficult for the trained model to detect difficult samples. Thus, we propose a new data augmentation method by simulating the irregular change process of the diseased tissue.

(4) To make more effective use of limited training data, we propose a dynamic update training method based on the sample complexity. Moreover, to further improve the generalization ability of the model, we also adopted a transfer learning method to assist the model training.

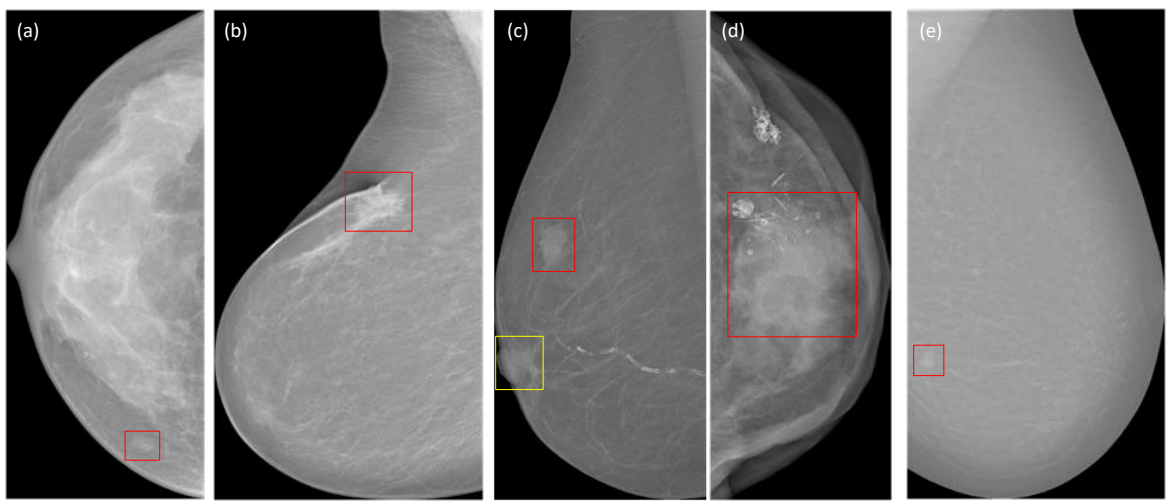

Fig. 1. Several typical examples of BMass, where the red box indicates BMass lesions and the yellow box indicates non-BMass tissues.

## 2  Related work

As early as the end of the 1960s, relevant personnel researched the detection and diagnosis of breast cancer [12]. With the development of medical image digitization technology and the continuous improvement of computer performance, from the early 1990s, academic and business circles have set off a research climax of computer-aided detection and diagnosis technology for breast cancer [13,14]. We roughly divide these breast mass detection methods into two categories, one is the traditional detection method, and the other is the deep learning-based detection method.

Although the traditional BMass detection method has certain limitations, it has made some progress [15–18]. For example, based on threshold segmentation [14], region growing [19,20], watershed [20,21], edge detection [22,23]. Specifically, Dubey R B et al. used a horizontal image set and a watershed two traditional image segmentation algorithms to segment the BMass, achieving an accuracy of 82% [21]. Xu et al. first preprocessed the mammography image, then extracted the boundary using the canny operator, and finally used the dynamic contour algorithm to post-process the boundary. Although the detection effect has improved, the time consumption is more obvious [22]. De Sampaio W B et al., considering the difference between dense and non-dense breast images, an adaptive algorithm is proposed to distinguish them, and then the genetic algorithm is used to segment BMass [24].

With the rapid development of deep learning in recent years, many CNN-based methods have also achieved good results in breast mass detection [25–29]. For example, Ribli D et al. first segmented the breast from the mammography image and then used the two-stage object detection algorithm Faster RCNN for BMass detection [30]. Akselrod-Ballin A et al. first preprocessed the image, then divided the dataset into three groups, benign BMass, malignant BMass, and others, and then used three modified Faster RCNN models to detect them separately [31]. Carneiro N D G et al. considered the problem of the large difference in the size of breast mass, proposed a multi-scale deep belief network to perform coarse detection on BMass, and used the Gaussian mixture model to classify candidate regions to reduce false positives [32]. To adapt to the smaller BMass lesions in the breast, Jung H et al. used RetinaNet, which is excellent for small object detection on ImageNet, as the basic model. Then, through various data augmentation methods such as rotation, flipping, and trimming, the detection performance of the small BMass is improved [33].

The detection method used in this paper has the following differences from the previous methods. 1) Adopting the new detection architecture FSAF based on anchor-free; 2) Proposing a new normalization method and combining it with the adaptive histogram equalization method to improve the contrast between BMass and surrounding tissue; 3) For the BMass detection task, we propose a new data augmentation method; 4) To make better use of limited training data, an effective new training method is proposed.

## 3  Methods

The BMassDNet detection architecture proposed in this paper is divided into three modules, namely image preprocessing, data augmentation, and BMass detection. The workflow is shown in Fig. 2. In general, preprocessing of the original image is necessary because of the large amount of black background in the mammography image and the low contrast between the tissues in the breast. Then, the preprocessed image is sample-expanded by the proposed data augmentation method to solve the overfitting problem caused by the small dataset. Finally, the object detection method based on the FSAF architecture is used to detect the BMass to improve the robustness of the detection system.

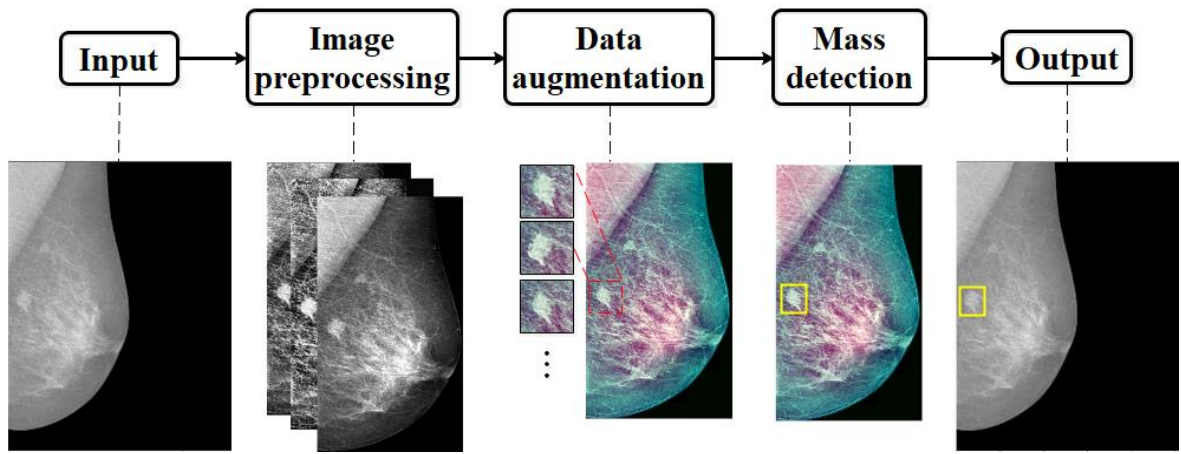

Fig. 2. Schematic diagram of the workflow for BMass detection.

## 3.1 Image preprocessing method

Fig. 3(a) and Fig. 3(b) show representative examples of the two datasets, INbreast and DDSM, respectively. The lower right corner is a partial enlarged image of the BMass region and a corresponding BMass mask image. Through observing the dataset, it can be found that there are many BMass lesions embedded in the glandular tissue or dense block of the breast, and BMass is very similar to these tissues, resulting in the BMass boundary being extremely unclear.

To solve the above problem, we first need to segment the breast and then normalize the rectangular region (ROI) where the breast is located by the proposed truncated normalization method. Thereafter, the ROI image is intensity stretched using adaptive histogram equalization to increase the contrast between BMass and other tissues. However, since adaptive histogram equalization is a nonlinear transformation method, it will change the original distribution of the image, which will have a certain impact on the learning of the model. To avoid this problem, we use the normalized and intensity stretched images together as input to the deep learning model.

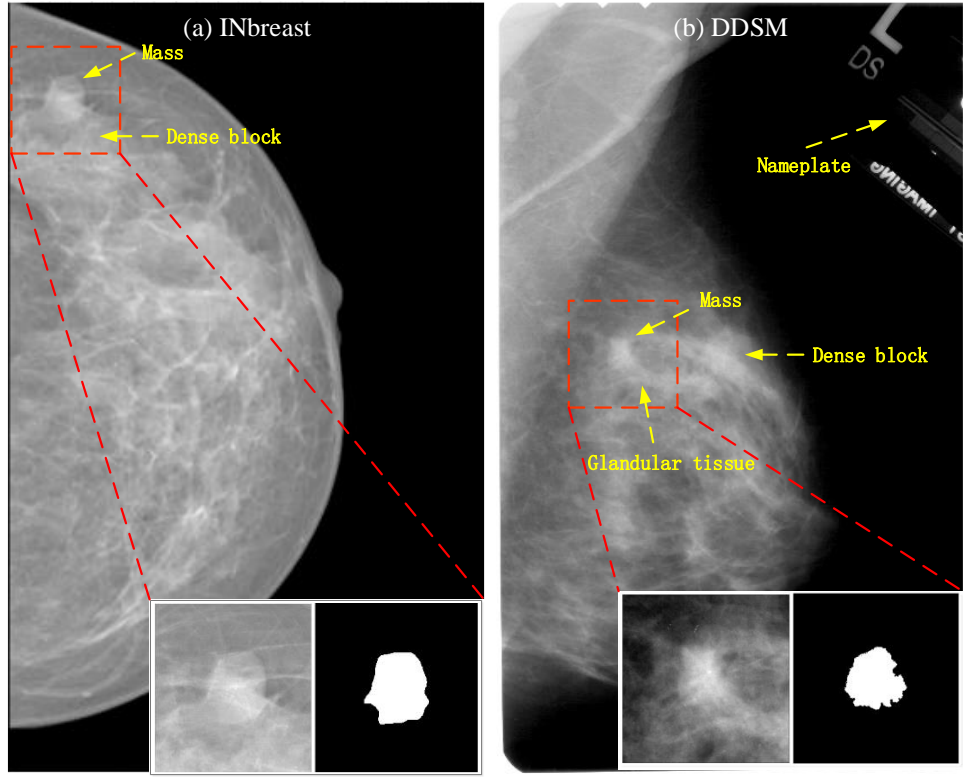

Fig. 3. Original mammographic images of INbreast and DDSM.

### 3.1.1 Segmentation of the breast

Since the boundary between the breast region and the black background region is very obvious in the mammography image, only a simple threshold method is enough to segment the breast. Specifically, we first perform Gaussian filtering on the original image; then segment the breast using the OTSU threshold method [34]. Finally, due to the existence of information such as nameplates, the image after OTSU segmentation may contain multiple connected regions, but since the area of the breast is generally the largest, we choose the largest connected region as the mask image of the breast.

### 3.1.2 Truncation normalization method

Although we removed most of the black background, the ROI image still has the problem of polarization of the intensity distribution. As shown in Fig. 4, the ROI image still has some black background, and the intensity of the truly effective breast region is compressed to less than one-third of the total distribution interval. Therefore, if the ROI image is directly normalized, it is likely to have an adverse effect on the detection of BMass.

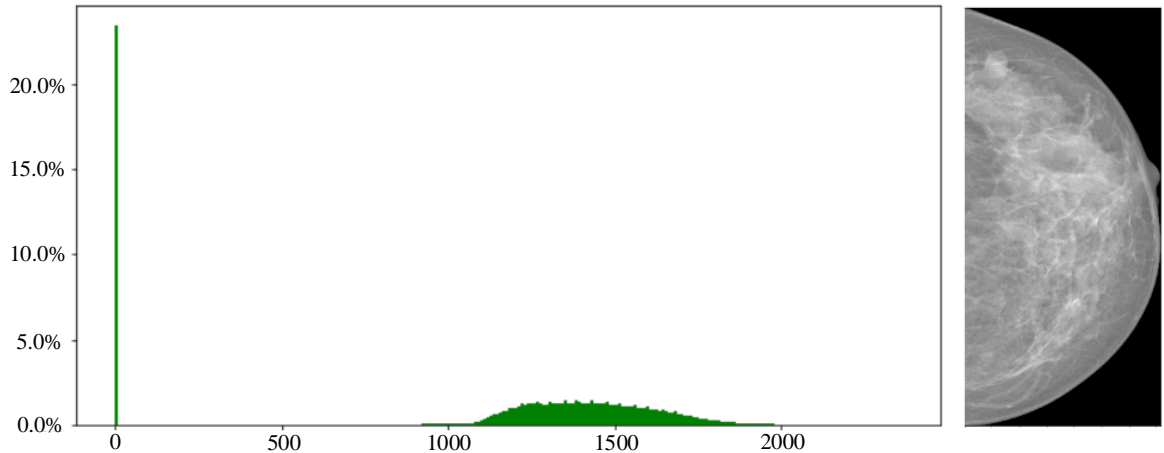
Fig. 4. Intensity histogram of ROI image.

To solve this problem, this paper proposes an adaptive normalization method called truncation normalization. Its basic idea is: according to the intensity histogram of the ROI image, select a pair of effective maximum intensity and minimum intensity, and then use them to cut off the intensity of the image, and then perform normalize operating. This will ensure that the breast region has a sufficient range of intensity distribution.

The specific steps of the method are as follows:

(1) Obtain the mask of the breast, and the method in Section 3.1.1 can be used directly here.

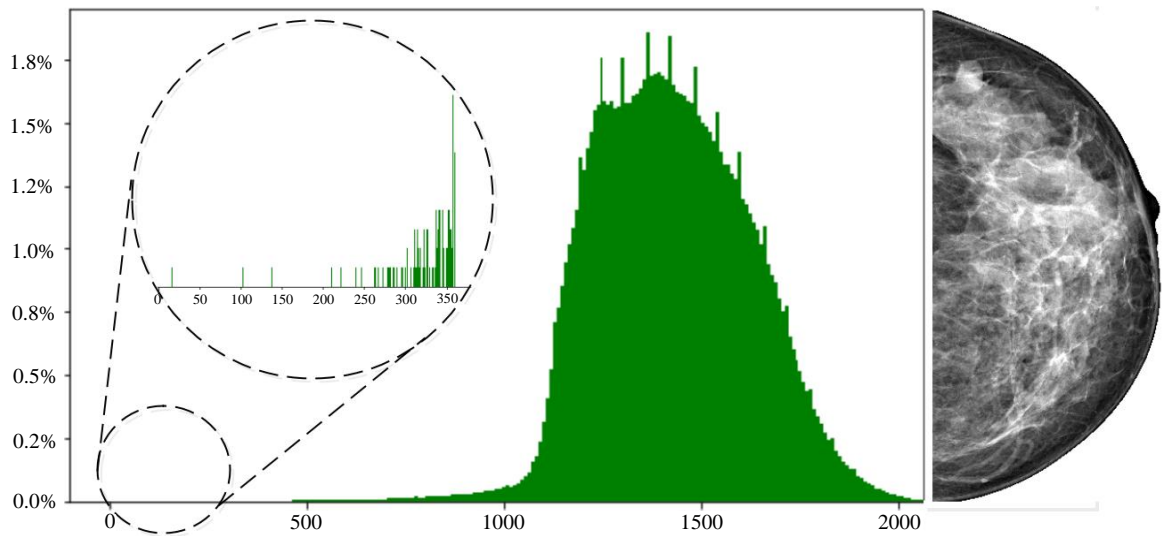
Fig. 5. The intensity distribution of a breast image without a background.

(2) Use the breast mask to extract the intensity image corresponding to the breast region, and then find the intensity distribution, as shown in Fig. 5. It can be seen that although we have removed all non-breast regions, there are still a few pixels with very small intensities. Moreover, the intensity range of the breast region still occupies less than one-half of the distribution interval.

(3) Sort the intensity of the breast, and then select 5% at the small end position as the minimum intensity Pmin, and select 1% of the large value end position as the maximum intensity Pmax. Then, the intensity P of each pixel in the image is subjected to the truncation processing as shown in Eq. (1). In this way, the influence of noise and a few abnormal values in the breast image can be avoided to a certain extent, and the original intensity distribution of the effective breast region is retained to the greatest extent.

$$P = \begin{cases} P\min, & if\ P \le P\min \\ P, & if\ P\min < P < P\max \\ P\max, & if\ P \ge P\max \end{cases} \quad (1)$$

(4) The intensity of each pixel of the truncated breast image is normalized, and the calculation formula is as shown in Eq. (2).

$$P = \frac{P - Pmin}{Pmax - Pmin} \quad (2)$$

Fig. 6 shows the effect of truncation normalization, in which Fig. 6(c) and Fig. 6(a) are the normalized image and its intensity distribution, respectively, and Fig. 6(b) is the original image. It can be seen from Fig. 6 that after the truncated normalization process, the intensity of the breast region has covered the entire interval, and the boundary of the BMass becomes clearer than the original image.

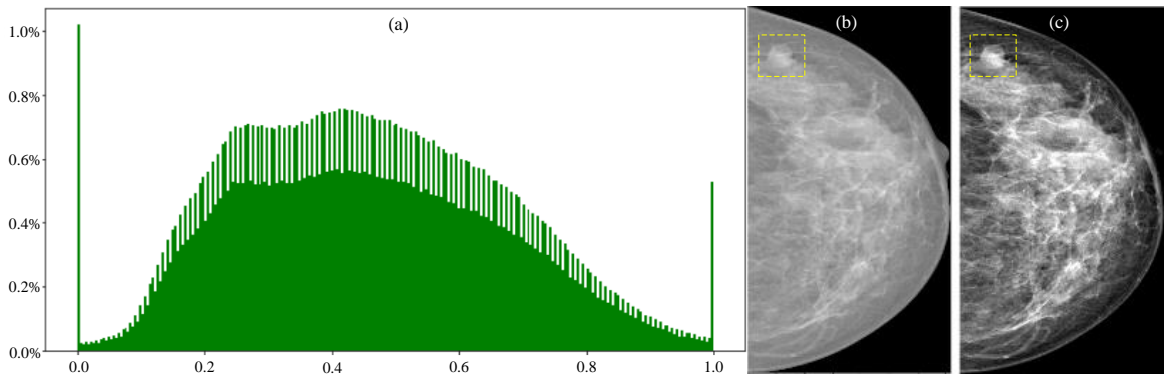

Fig. 6. The effect of truncating normalization. Among them, Fig. 6(c) and Fig. 6(a) are the normalized image and its intensity distribution, respectively, and Fig. 6(b) is the original image.

### 3.1.3 Image enhancement method

Both the traditional lesion detection method and the deep learning-based lesion detection method require significant differences between the object and the background, that is, the object has a clear boundary. Due to the adhesion of some of the BMass to the dense block or glandular tissue in the mammography image, their boundaries are very blurred. The main purpose of our prior use of truncated normalization is to preserve the intensity difference of the breast region to the greatest extent. It can only alleviate the problem of unclear boundaries to some extent. Therefore, in addition to the truncating normalization of ROI images, we also need to use more effective image enhancement methods to improve the contrast between BMass and other tissues.

Adaptive Histogram Equalization (AHE) algorithm can achieve good results in both natural image and medical image enhancement, but the AHE algorithm has such a problem. That is, when the intensity of a certain region is relatively uniform, the histogram of the region will have an intensity peak, and after the equalization, the intensity originally in a narrow interval will be stretched to the entire intensity range, thereby introducing noise. This is not what we expected. Therefore, this paper uses AHE's improved algorithm, Contrast Limited Adaptive Histogram Equalization (CLAHE) algorithm [35] to enhance ROI images.

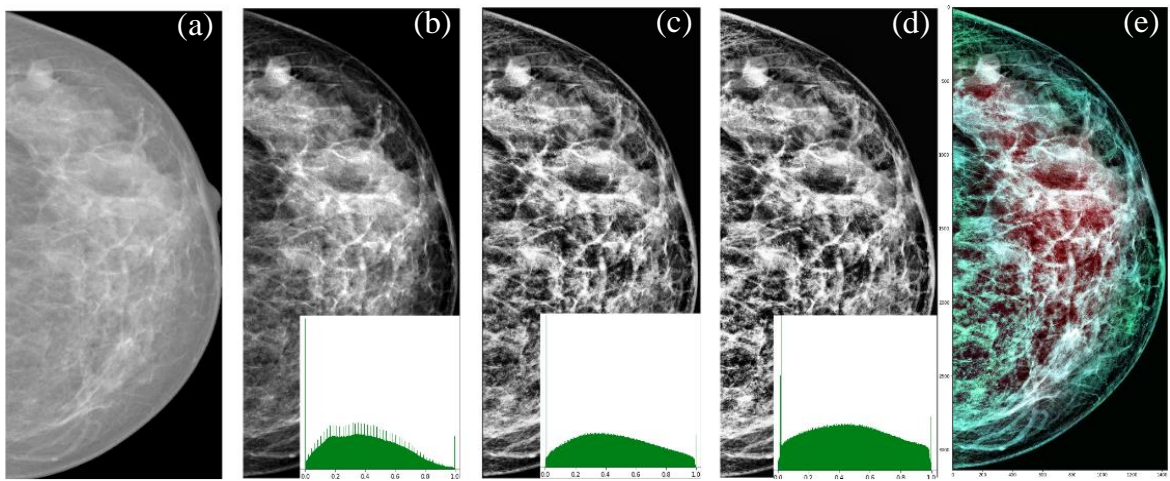

Fig. 7. The comparison of the original image, the truncated normalized image, and the enhanced image. The lower right corner of the image is the intensity histogram corresponding to the breast region, where (a) represents the original image; (b) represents the image after truncation normalization; (c, d) represents the enhanced image when the clip-limit is 0.01 and 0.02, respectively. (e) represents the synthesized images of (b), (c) and (d).

Fig. 7 shows the original image, the truncated normalized image, and the enhanced image, respectively. It can be seen that the BMass boundary is very blurred in the original image, especially the lower side of BMass shows almost an inseparable state with the glandular tissue. The truncated normalized image improves the problem to some extent. The upper boundary of BMass becomes clearer, but the lower boundary is still blurred. The clarity of the boundary of the BMass enhanced by CLAHE has been further improved. To ensure that the model can fully extract effective features, we use (b), (c), and (d) as three channels of the color image to synthesize the color image shown in (e) and use it as the input image of the deep learning model.

## 3.2 Natural deformation data augmentation method

Data is the key to deep learning models, that is, the performance of deep learning models is largely dependent on the size and diversity of the dataset. Medical images have very limited access to data compared to natural images. Moreover, the label needs to be labelled by a professional radiologist, and even a biopsy is needed to give an accurate label, so the size of the medical image dataset is generally small. In particular, the distribution of lesions in medical images is sparse, resulting in an imbalance in the number of positive and negative samples in the data set. Therefore, increasing the number and diversity of positive samples has become a key to deep learning models in medical imaging applications.

| **Algorithm 1: natural deformation data augmentation method** | |
|---|---|
| **Input:** | Original image and mask image of BMass |
| **Output:** | The deformed image. |
| **Step1:** | The original image is decomposed into a BMass image and a background image according to the BMass mask. |
| **Step2:** | The same random deformation is performed on the BMSs image and the background image, respectively, using the elastic deformation algorithm [36]. |
| **Step3:** | The deformed BMass image is filled into the original background image, and the mask of the residual region is calculated. |
| **Step4:** | The corresponding region image (ResImg) is extracted from the deformed background image according to the mask of the residual region, and ResImg is filled to the corresponding position of the output image of Step3. |
| **Step5:** | Use the stain filling algorithm [37] to repair the BMass edge to get the output image. |

Traditional data augmentation methods include random rotation, random translation, mirroring, and random clipping. Such methods can expand the sample size of the dataset to a certain extent, but cannot alleviate the problem of sparse positive samples in medical images. Therefore, this paper proposes a data augmentation method based on local elastic deformation, which we call the "natural deformation data augmentation method". The main idea is that for an image containing BMass, only the BMass is elastically deformed to simulate the natural change of BMass, and the local background region in contact with BMass also changes. The algorithm flow of natural deformation is shown in Algorithm 1.

As shown in Fig. 8, the processing results of the respective steps of Algorithm 1 are given. Since the resolution of the mammography image is high, only the image of the region near the BMass is shown here.

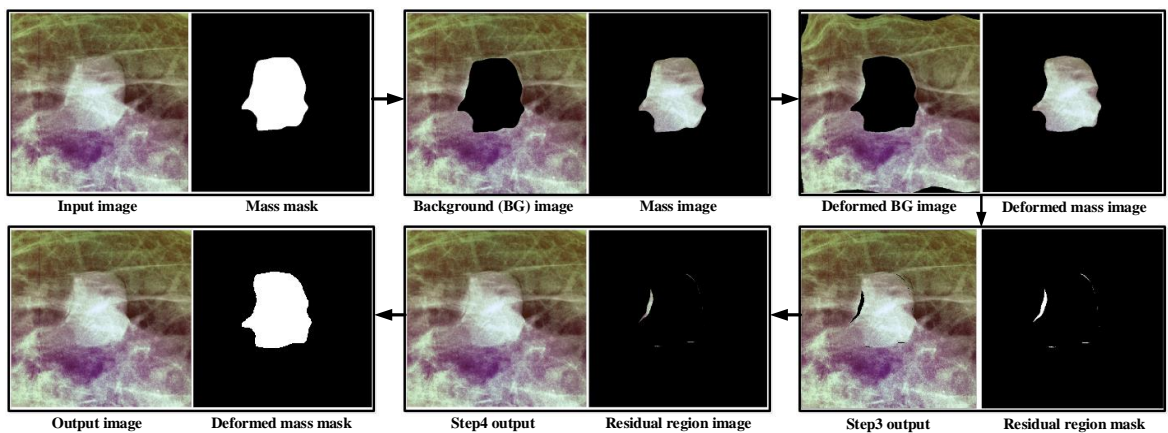

Fig. 8. Results for each step in Algorithm 1.

Fig. 9 shows the effect of natural deformation of several images randomly selected from the dataset, where O1-O6 is the original image and D1-D7 is the natural deformed image. It can be seen from Fig. 9 that the image that has undergone natural deformation is relatively natural, and there is no obvious trace of deformation, which achieves the intended purpose.

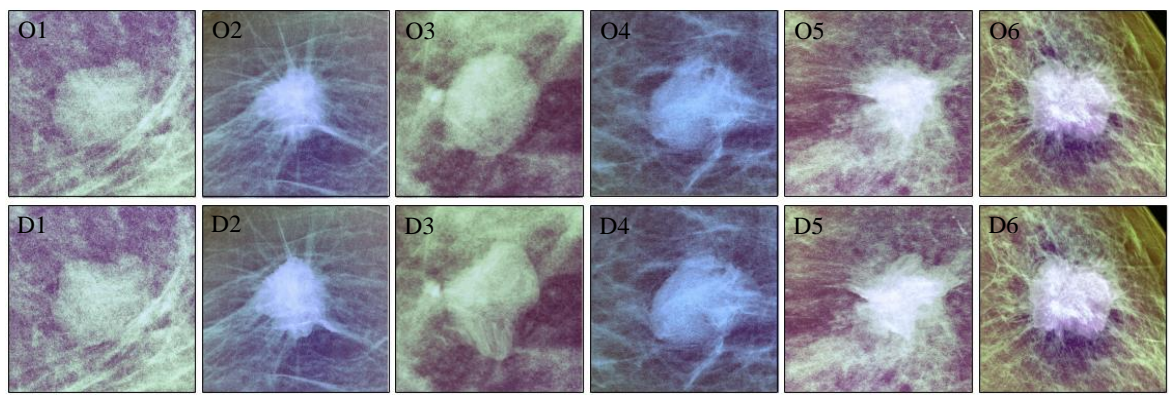

Fig. 9. Visual representation of natural deformation.

Though the subtle change in such local deformation is difficult to discern by the human eye, it may be easily recognized by the computer. In the process of data augmentation, we not only apply Algorithm 1 to the region where BMass is located but we also randomly apply it to other regions on the breast. In this way, during the continuous learning process, the machine will think that the change caused by this deformation is inherent to the real sample, not artificial, which will indirectly make the expanded sample more natural and real. Also, we will also verify the necessity of this practice through experiments. See Section 4.4 for specific experiments.

### 3.3   Detection of breast mass

At present, the methods of deep learning in the field of object detection are mainly divided into two categories: two-stage and one-stage. The two-stage object detection method first uses a region proposal network (RPN) to extract the suspicious target region and then extracts the ROI corresponding to the suspicious region from the feature map for classification and regression. The one-stage object detection method does not use RPN to extract the ROI of the suspect region, but each feature value on the original feature map is classified and regressed. Among them, the range of images that can be seen by each feature value is determined by the receptive field of the model.

The BMass detection based on mammography images studied in this paper is different from the target detection in natural images. Specifically, in the natural image, the object and the background are independent of each other, and the carcinogenesis of BMass in the mammography image is bound to cause lesions in the surrounding tissue. That is, the background information around the BMass can provide additional features to help the model categorize more accurately. If a two-stage detection architecture is used, the ROI extracted by the first stage RPN will limit the effective range of the second stage feature extraction and will lose the spatial position information of BMass in the breast. This makes it difficult to distinguish between lymph nodes, dense blocks, and the like that are morphologically similar to BMass. Therefore, we believe that the one-stage object detection method is more suitable for BMass detection tasks.

### 3.3.1   Micro-modified FASF architecture

There are two difficulties in using the convolutional neural network for object detection. Firstly, the classification of objects requires abstract deep features, but the location information of deep features is not accurate. Secondly, the size of the object in the image varies. This paper uses the one-stage object detection architecture FSAF with a feature pyramid network (FPN) structure to solve the above problems. The FSAF [11] detection architecture proposed by Zhu et al. is an improved version of the current popular RetinaNet [38] detection architecture. It is mainly composed of feature extraction network, FPN, anchor-based classification regression network, and anchor-free feature selection module.

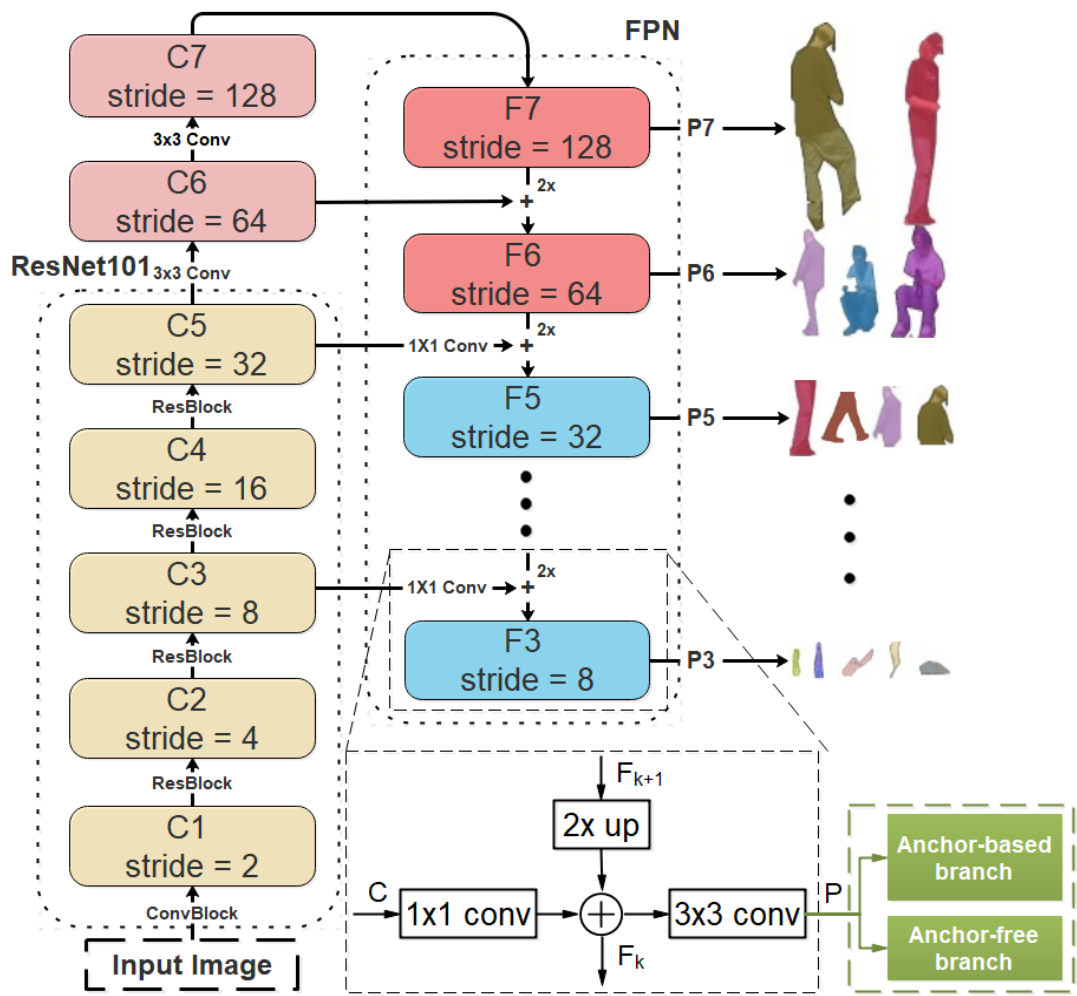

Fig. 10. Network architecture diagram of the micro-modified FSAF detection model

Fig. 10 shows a network architecture diagram of the FSAF detection model used herein. Among them, {C1~C7} represents the output feature map of each convolution operation in the feature extraction network; {F3~F7} represents the feature map in the FPN; {P3~P7} represents the input of the anchor-based branch and the anchor-free branch. Different from the original FSAF detection architecture, we did not directly output {C6, C5}, but generated {P6, P5} according to the generation method of {P3~P4}. That is, we combine low-level features and high-level features to better detect objects using horizontal connection and upsampling.

### 3.3.2 Dynamic update training method

Due to the limited number of samples in the training set, it is difficult to train a robust detection model. Therefore, we use the idea of transfer learning to alleviate this problem. In other words, this paper uses the parameters trained on ImageNet by the ResNet101 classifier as the initialization parameters of the FSAF feature extraction network. Among them, the FPN and classification regression networks of the FSAF detection architecture do not have pre-training weights and need to be trained from the beginning.

Besides, to make better use of limited training data, we need an efficient training method. We know that the validation set is generally used to reflect the performance of the model during training, so as to guide the setting of the relevant hyper parameters, it is not directly involved in the training of the model. For the medical image field, where data size is scarce, this approach is very extravagant. To solve the above problems, Jiao Z et al. proposed a dynamic update training strategy [39]. Its main idea is to add the samples of the detection errors (hard samples) in the validation set to the training set, and randomly select the same number of samples from the training set into the validation set, and then iterate until the performance of the validation set is no longer improved. However, there is a problem with this strategy, that is, when updating the dataset, all the hard samples are added to the training set, resulting in too many hard samples in the training set, and the validation set is almost all simple samples. This makes the validation set not well reflect the performance of the model, resulting in instability of the training process.

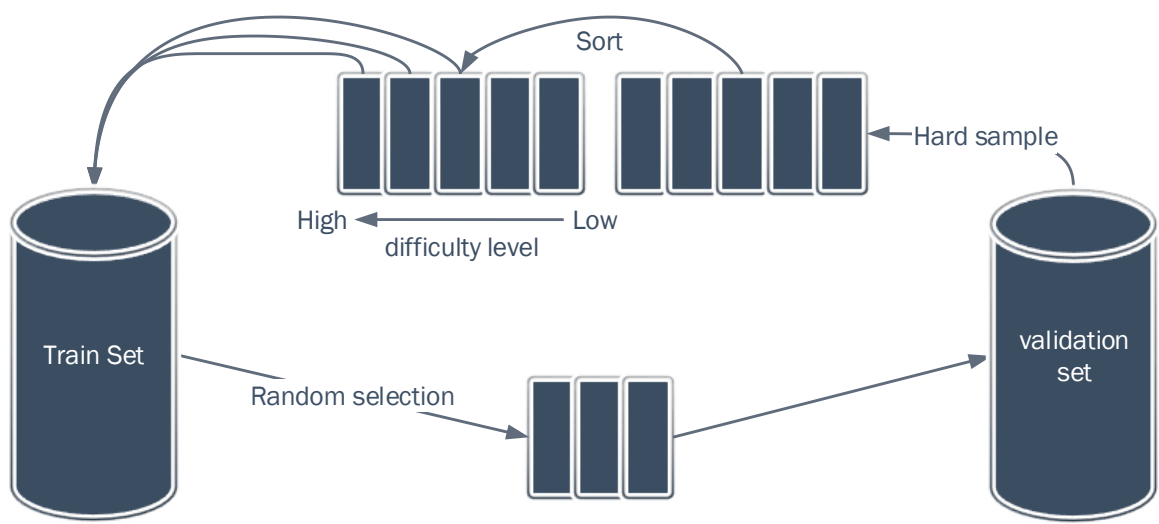

Fig. 11. Schematic diagram of the work of the dynamic update training method based on the difficulty level of the sample.

Given the above problems, this paper proposes a dynamic update training method based on the difficulty level of the sample. The basic idea is that, at each iteration, the hard samples in the validation set are sorted according to the degree of difficulty, and then a certain number of hard samples are selected and exchanged with the samples in the training set. Fig. 11 shows a workflow diagram of the proposed training method.

As shown in Fig. 11, in the training process of the model, the training data is first divided into a training set and a validation set according to a certain ratio, and then iterative training is started. Then, according to the loss value of the hard samples, the values are sorted from small to large, and the three samples with the highest loss value are selected to join the training set, and three samples are randomly selected from the training set to be added to the validation set. The condition for the end of the iteration is to stop the update of the data when no hard samples are found in the validation set. Finally, all the samples in the validation set are added to the training set, and the training will continue for ten epochs with a smaller learning rate (10% of the initial learning rate) to complete the final training of the model.

## 4 Experiment and results

Below we will explain the data used, evaluation criteria, and experimental details.

### 4.1 Data

The experiments carried out in this paper are mainly based on the high-quality public dataset INbreast [40]. The INbreast dataset is from the Porto Breast Central Hospital in Portugal and contains 115 female mammography images. Since a breast generally corresponds to two views, an MLO (mediolateral oblique) view, and a CC (craniocaudal) view, the theoretical INbreast dataset should contain 460 images. But since 25 of the 115 women had undergone mastectomy, which resulted in only two images for these individuals, the INbreast dataset had only 410 images. A total of 107 images of the 410 images contained BMass lesions, of which 35 were BMass malignant lesions, and 72 were BMass benign lesions. Also, to verify the robustness of the model, we conducted related comparison experiments on the CBIS-DDSM dataset [41]. Finally, we use a 2-fold cross-validation method to evaluate the performance of the model based on the number of views for each breast.

## 4.2 Evaluation criteria

To quantitatively measure the performance of the proposed BMass detection method, the true positive rate (TPR) and the number of the false positive per image (FPPI) were used as the evaluation indicators of the detection model.

For the detection task of BMass, the predicted value of the model is the bounding box and confidence (Conf) of all suspicious lesions, the bounding box indicates the position of the suspected lesion in the image, and the confidence indicates the probability that the lesion belongs to BMass. In addition, this paper determines whether the prediction box given by the model can be accepted according to the judgment criteria given in Table 1. Among them, IOU (Intersection Over Union) indicates the accuracy of detection, Conf_TH indicates the minimum value of Conf when the model considers that the current lesion belongs to BMass, and IOU_TH indicates the minimum value of IOU when the model considers that the detection is correct. Note that the IOU_TH is fixed at 0.5 in the experiment. If the Conf of the prediction box is greater than or equal to Conf_TH, and the IOU of the prediction box and the GT (Ground True) is greater than or equal to IOU_TH, the prediction box is determined to be true positive (TP). Similarly, FP indicates a false positive, FN indicates a false negative, and TN indicates a true negative.

Table 1. Decision table for the correctness of the prediction box.

|  | IOU ≥ IOU_TH | IOU < IOU_TH |
| --- | --- | --- |
| Conf ≥ Conf_TH | TP | FP |
| Conf < Conf_TH | FN | TN |

The calculation formulas for TPR and FPPI are as shown in equations (3) and (4).

$$TPR = \frac{TP}{TP + FN} \tag{3}$$

$$FPPI = \frac{FP}{Number\ of\ Images} \tag{4}$$

## 4.3 Experimental details

The detection task of the lesion in the medical image generally requires the model to have a high recall rate. Therefore, in the training of the BMass detection model, only the image containing BMass is selected. To alleviate the model over-fitting problem, in addition to using the proposed natural deformation data augmentation method, we also use six traditional data augmentation methods such as random rotation (-0.1º~0.1º), random translation(0~0.1), random shear (-0.1º~0.1º), random scaling (0.9~1.1), horizontal flip, and vertical flip.

Since FSAF is a fully convolutional neural network, the size of the input image is not fixed. In training and testing, we first scale the shorter side of the image to 800 pixels; then scale the long side to the corresponding size according to the aspect ratio of the original image, while limiting the size of the long side to no more than 1333 pixels. The other hyper parameters during training are set, the total number of training epoch is 50; since the size of the input data varies, the batch size is set to 1; the initial learning rate is 1e-5, and if the loss values of two consecutive generations are not decreased, the learning rate is reduced to the original 10%; Adam optimizer has a gradient cutoff of 0.001.

## 4.4 Overall performance

To more intuitively observe the overall performance of the proposed breast mass detection method, as shown in Fig. 12, we plotted the FROC curve corresponding to BMassDNet on the INbreast and DDSM datasets.

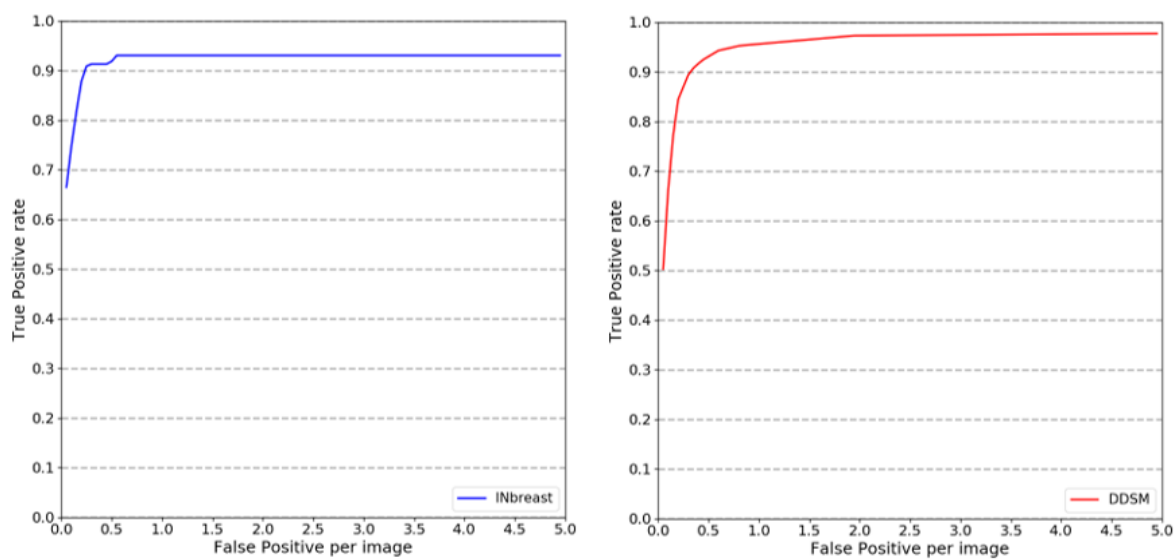

Fig. 12. FROC curve of BMassDNet on INbreast and DDSM.

By observing Fig. 12, it is found that when the number of false positives of each image is close to 0, the performance of BMassDNet on the INbreast dataset is better than DDSM.

When the number of false positives per image exceeds 0.25, the true positive rate of BMassDNet on both datasets reaches 90%. Besides, with the increasing number of false positives per image, the performance of BMassDNet on INbreast has almost stabilized, but the performance on DDSM is still slowly improving. Overall, the breast mass detection network based on the anchor-free architecture can be adapted to the breast mass detection task to a certain extent.

# 5 Discussion

Below we will explain the ablation study, and an empirical comparison.

## 5.1 Ablation study

To verify the effectiveness of the components in the BMassDNet detection architecture proposed in this paper, we designed the ablation experiments as shown in Table 2. Table 2 shows the results of removing only one component on the final used BMassDNet architecture. Overall, the data comparison of the first eight rows (except the header) in Table 2 verifies the effectiveness of the components in the BMassDNet architecture. The comparison of the data in the last four rows of Table 2 can intuitively explain the reason for choosing FSAF instead of using other one-stage object detection networks. The details are as follows.

(1) The effectiveness of the TNM

By comparing the first two sets of data in Table 2, it can be found that after removing the TNM, the TPR is reduced by nearly one point, and the FPPI is also increased by nearly four times. This result shows that we validate the effectiveness of the proposed truncation normalization method from an experimental point of view.

(2) The effectiveness of the IEM

Comparing the data of the first and third sets of Table 2, it can be seen that after removing the IEM, the TPR is reduced by 0.9% and the FPPI is increased by 0.42. That is, the overall performance of the model has decreased, which illustrates the necessity of intensity stretching of the original image.

(3) The effectiveness of the NDDA

Comparing the data of the first and fourth groups of Table 2, it can be seen that after removing the NDDA-expanded sample, the TPR is significantly reduced and the FPPI is greatly increased, which fully reflects the effectiveness of the NDDA. That is, it can significantly improve the generalization performance of the model.

(4) The effectiveness of the NDDA_NMR

By comparing the data from the first and fifth sets, it can be found that although the natural deformation in the non-BMass region has little effect on the TPR, the FPPI is reduced by 0.493. This confirms our analysis in Section 3.2 that natural deformation in the non-BMass region facilitates model learning. It should be noted that when Algorithm 1 is executed in the non-BMass region, the input mask is a local region that needs to be specified for natural deformation.

Table 2. Ablation experiment. Note that, "TNM" indicates the truncation normalization method; "IEM" indicates image enhancement method; "NDDA" indicates the natural deformation data augmentation method; "NDDA_NMR" indicates that natural deformation is performed only in the local region where BMass is located, and natural deformation is not performed in the non-mass region.; "DUTM" indicates the dynamic update training method; "PreWeight" indicates the weight of training on ImageNet during transfer learning; "NMS" represents the non-maximum suppression.

| Methods | TPR | FPPI |
|---|---|---|
| BMassDNet | 0.913 | 0.256 |
| BMassDNet – TNM | 0.904 | 1.009 |
| BMassDNet – IEM | 0.904 | 0.676 |
| BMassDNet – NDDA | 0.861 | 1.493 |
| BMassDNet – NDDA_NMR | 0.913 | 0.749 |
| BMassDNet – DUTM | 0.887 | 0.466 |
| BMassDNet – PreWeight | 0.652 | 3.546 |
| FSAF [11] | 0.913 | 0.512 |
| YOLO [42] | 0.800 | 0.358 |
| RetinaNet [38] | 0.913 | 0.524 |
| FCOS [43] | 0.904 | 0.480 |

(5) The effectiveness of the DUTM

Comparing the data of the first and sixth groups, the proposed DUTM not only increases the TPR by 2.6% but also reduces the FPPI by 0.21, which is a good proof of the effectiveness of the proposed training method.

(6) The effectiveness of the PreWeight

Comparing the data of the first and seventh groups, it can be seen that loading the weights already trained on ImageNet can greatly improve the performance of the detection model. Specifically, the TPR increased by 0.261 and the FPPI decreased by 3.29.

(7) The effectiveness of the Micro-modified FASF

Comparing the data of the first and eighth sets, we can see that our modification of the original FASF architecture is beneficial to the detection of BMass. Although the TPR has not improved, the FPPI has been reduced by half, which explains to some extent that our revision of the FSAF architecture is correct.

(8) The reason for choosing the FSAF structure.

In Section 3.3, we explain why we chose a one-stage object detection network. Here, we will explain the reasons for choosing the FSAF architecture through experimental comparison. Specifically, in addition to the FSAF architecture, we have implemented three popular one-stage object detection architectures, namely YOLO, RetinaNet and FCOS. By comparing the last four lines of data, the detection based on the FSAF architecture is the best. Although YOLO has the lowest FPPI, its corresponding TPR is also relatively low, which is not desirable in medical applications. For the sake of fairness, the last four lines are identical except for the network architecture.

## 5.2  Experimental comparison

To verify the superiority of the proposed BMassDNet, we compared it with the BMass detection method published in recent years. Table 3 shows the experimental comparison of BMassDNet on the two datasets of INbreast and DDSM. It is not difficult to see from Table 3 that since the INbreast data set is publicly available in 2012, and deep learning is beginning to flourish at this time, the BMass detection method developed on INbreast is rarely based on traditional methods. By comparing the experiments on the INbreast dataset, we have proposed the BMassDNet method, which shows excellent performance regardless of whether the sample being evaluated contains normal samples. Here, we will further verify by experiment, for the detection of BMass, the one-stage detection method is better than the two-stage detection method. Specifically, the methods proposed by Ribli et al. and Akselrod-Ballin et al. are based on the most typical two-stage detection method, Faster RCNN, but the performance of BMassDNet is somewhat superior to them. That is, in the case where the Recall is almost the same, the FPPI is five points lower than them.

Table 3. A comparison on the quantitative results of various mass detection methods. Please note that, "TM" indicates the traditional method, "DL" indicates the deep learning.

| Datasets | Types | Methods | Images | TPR | FPPI |
|---|---|---|---|---|---|
| INbreast | DL | Amit et al. [25] | 107 | 0.870 | 1.423 |
| | DL | Dhungel et al. [32] | 410 | 0.870 | 0.800 |
| | DL | Wu et al. [26] | 107 | 0.880 | 0.750 |
| | DL | Ribli et al. [30] | 410 | 0.900 | 0.300 |
| | DL | Jung et al. [33] | 410 | 0.910 | 1.300 |
| | DL | Agarwal et al. [27] | 410 | 0.920 | 2.594 |
| | DL | Akselrod-Ballin et al. [31] | 100 | 0.930 | 0.560 |
| | DL | Agarwal et al.[29] | 410 | 0.870 | 0.250 |
| | DL | **BMassDNet (Conf=0.2)** | **107** | **0.930** | **0.495** |
| | DL | **BMassDNet (Conf=0.6)** | **410** | **0.913** | **0.256** |
| DDSM | TM | Tai et al. [15] | 358 | 0.900 | 4.800 |
| | TM | Sampaio et al. [24] | 678 | 0.837 | 0.190 |
| | TM | Nazaré Silva et al. [16] | 599 | 0.923 | 1.120 |
| | TM | Anitha et al.[17] | 300 | 0.925 | 1.060 |
| | TM | Vikhe et al. [18] | 90 | 0.939 | 0.740 |
| | DL | Bandeira Diniz et al. [28] | 266 | 0.904 | 0.878 |
| | DL | **BMassDNet (Conf=0.3)** | **890** | **0.943** | **0.599** |
| | DL | **BMassDNet (Conf=0.6)** | **890** | **0.843** | **0.197** |

However, since the DDSM dataset is publicly available for very early time, most of the BMass detection methods developed on it are based on traditional methods. Experiments conducted on the DDSM dataset show that although the traditional BMass detection method can have a better performance in a carefully selected dataset, its generalization ability is poor. In addition, since the traditional method requires artificial design of features, the limitations thereof are large. The deep learning-based approach does not have these limitations, and it automatically learns the relevant features. For example, the CNN-based BMass detection method proposed by Bandeira Diniz et al. can achieve good performance.

The detection performance of FSAF-based BMassDNet proposed in this paper has been further improved. Specifically, BMassDNet is superior to the method proposed by Vikhe et al., although the TPR is not much improved, the FPPI is reduced by 0.141. Also, when Conf took 0.6, BMassDNet also achieved similar performance to Sampaio et al.

# 6 Conclusion

In this study, we propose a BMass detection method based on FSAF network architecture, which not only can lighten the visual fatigue for physicians but also can raise the detection accuracy of BMass. Specifically, due to the large number of unrelated backgrounds in the mammography image and the blurring of the boundaries of the BMass, this poses certain difficulties for the training of the model. To solve this problem, we use the proposed truncation normalization method to preprocess the image and combine it with the adaptive histogram equalization algorithm to improve the contrast of the image. This ensures that the BMass has a clear boundary and maintains the data distribution of the original image. In addition, to alleviate the over-fitting problem caused by the small scale of the dataset, this paper proposes a natural deformation data augmentation method. This method can simulate the various forms of BMass irregular growth in the breast, so as to increase the diversity of training data. Meanwhile, in order to effectively use limited data, we propose a dynamic update training method based on the degree of sample difficulty. This training method makes our training process more stable and can more effectively mine hard samples. In terms of network structure, we have made appropriate improvements to the FSAF detection architecture, that is, we also use the last two layers of the feature extraction network for combination with low-level features in order to better adapt to the problem of different sizes of BMass. Finally, from the experimental point of view, through the ablation study and experimental comparison, the various components and overall performance of the proposed BMass detection method are verified. According to the experimental results shown in Table 3, we obtained the optimal performance compared to other existing methods.

It should be pointed out that although the method proposed in this paper has achieved good results, there are still some shortcomings. For example, the proposed natural deformation data augmentation method is slower to calculate and needs further optimization to speed up its operation.

## Conflict of interests

All authors declare that they have no conflicts of interest regarding the publication of this paper.

## Acknowledgements

The National Key R&D Program of China (Grant No. 2017YFC0112804) supported this work.